\documentclass[prd,twocolumn,groupedaddress,showpacs,nofootinbib]{revtex4}
\usepackage{graphicx}
\usepackage{dcolumn}
\usepackage{amssymb}
\usepackage{mathrsfs}
\usepackage{amsmath}
\usepackage{epsfig}
\usepackage[dvips]{color}
\usepackage{hhline}

\begin{document}

\title{Double type-II seesaw accompanied by Dirac fermionic dark matter }

\author{Pei-Hong Gu}
\email{peihong.gu@sjtu.edu.cn}

\affiliation{School of Physics and Astronomy, Shanghai Jiao Tong University, 800 Dongchuan Road, Shanghai 200240, China}

\begin{abstract}

In the type-II seesaw mechanism, the neutrino mass generation could be tested experimentally if the Higgs triplet is at the TeV scale and has a small cubic coupling to the standard model Higgs doublet. We show such small triplet-doublet coupling and the cosmic baryon asymmetry can be simultaneously induced by an additional seesaw mechanism involving a $U(1)_{B-L}^{}$ gauge symmetry. Meanwhile, three neutral fermions for cancelling the gauge anomalies can form a stable Dirac fermionic dark matter besides an acceptably massless fermion.

\end{abstract}


\maketitle

\section{Introduction}

The atmospheric, solar, accelerator and reactor neutrino experiments have established the phenomenon of neutrino oscillations. Three flavors of neutrinos thus should be massive and mixed \cite{pdg2018}. Moreover, the neutrinos should be extremely light to fulfil the cosmological observations \cite{pdg2018}. The tiny but nonzero neutrino masses call for new physics beyond the standard model (SM). Currently, the best explanation for the small neutrino masses seems to be the famous seesaw \cite{minkowski1977} mechanism. In some seesaw models \cite{minkowski1977,mw1980}, the interactions for generating the small neutrino masses can also produce a lepton asymmetry stored in the SM leptons \cite{fy1986,lpy1986,fps1995,ms1998,bcst1999,hambye2001,di2002,gnrrs2003,hs2004,bbp2005}. The sphaleron \cite{krs1985} processes then can partially convert the produced lepton asymmetry to a baryon asymmetry. This means the cosmic baryon asymmetry, which is another big challenge to the SM, can have a common origin with the small neutrino masses.

The type-II seesaw  \cite{mw1980} has become one of the  most attractive seesaw scenarios. In the type-II seesaw models, the Higgs triplet has a cubic coupling to the SM Higgs doublet. In the presence of a small triplet-doublet cubic coupling, the Higgs triplet can pick up a tiny vacuum expectation value (VEV) even if it is at the TeV scale. Accordingly, the Yukawa couplings of this TeV-scale Higgs triplet to the SM lepton doublets can be allowed at a testable level \cite{fhhlw2008}. Obviously, the key of the testable type-II seesaw is the small cubic coupling between the Higgs triplet and doublet. In an early work \cite{ghsz2009}, we introduced a global symmetry of lepton number to simultaneously explain such small triplet-doublet coupling and the cosmic baryon asymmetry. However, the spontaneous breaking scale of the global lepton number is quite arbitrary.

In this paper we shall realize a double type-II seesaw mechanism by resorting to a $U(1)_{B-L}^{}$ gauge symmetry which forbids the Yukawa couplings of three right-handed neutrinos to the SM. Through their Yukawa couplings to a Higgs singlet for spontaneously breaking the $U(1)_{B-L}^{}$ symmetry, the right-handed neutrinos eventually can form a Dirac fermion to become a stable dark matter particle besides a harmlessly massless state. Because of this $U(1)_{B-L}^{}$ symmetry breaking, two or more heavy Higgs singlets can acquire their small vacuum expectation values (VEVs) to suppress the cubic coupling between the usual type-II seesaw Higgs triplet and the SM Higgs doublet. Therefore, the left-handed neutrinos can naturally obtain their tiny Majorana masses even if the Higgs triplet with sizable Yukawa couplings is set at the TeV scale. Our model can also accommodate a successful leptogenesis mechanism through the heavy Higgs singlet decays.

\section{Fermions and scalars}

 The SM fermions and scalar are denoted as follows, 
\begin{eqnarray}
&&\begin{array}{l}q^{}_{L}(3,2,+\frac{1}{6},+\frac{1}{3}),\end{array} \!\!
\begin{array}{l}d^{}_{R}(3,1,-\frac{1}{3},+\frac{1}{3}),\end{array}  \!\!
 \begin{array}{l}u^{}_{R}(3,1,+\frac{2}{3},+\frac{1}{3});\end{array} \nonumber\\
[2mm]
&&\begin{array}{l}l^{}_{L}(1,2,-\frac{1}{2},-1),\end{array} \!\!
\begin{array}{l}e^{}_{R}(1,1,-1,-1);\end{array} \!\! \begin{array}{l}\phi^{}(1,2,-\frac{1}{2},0).\end{array}
\end{eqnarray}
Here and thereafter the brackets following the fields describe the transformations under the $SU(3)_c^{} \times SU(2)^{}_{L}\times U(1)_Y^{}\times U(1)_{B-L}^{}$ gauge groups. For simplicity, we do not show the indices of the three generations of fermions. In order to cancel the gauge anomalies, we need some right-handed neutrinos with appropriate $U(1)_{B-L}^{}$ charges. In the present work, we shall consider the following three right-handed neutrinos \cite{mp2007},
\begin{eqnarray}
\begin{array}{l}
\nu^{}_{R1}\left(1,1,0,-4\right),\end{array}\!\!  \begin{array}{l}\nu^{}_{R2}\left(1,1,0,-4\right),\end{array}\!\! \begin{array}{l}\nu^{}_{R3}\left(1,1,0,+5\right).\end{array}
\end{eqnarray}

The Higgs doublet $\phi$ is responsible for the electroweak symmetry breaking as usual. We then introduce a Higgs singlet, 
\begin{eqnarray}
\begin{array}{l}
\xi(1,1,0,+1),\end{array}
\end{eqnarray}
for spontaneously breaking the $U(1)_{B-L}^{}$ symmetry. Moreover, the model contains other Higgs scalars including a Higgs triplet, 
\begin{eqnarray}
\begin{array}{l}
\Delta(1,3,+1,+2),\end{array}
\end{eqnarray}
and two or more heavy Higgs singlets,
\begin{eqnarray}
\begin{array}{l}
\sigma^{}_{a}(1,1,0,-2),~~(a=1,...,n\geq 2).\end{array}
\end{eqnarray}

For simplicity, we do not write down the full Lagrangian. Instead, we only give the following terms,
\begin{eqnarray}
\label{lag}
\mathcal{L}&\supset& -\lambda_{\xi\phi}^{}\xi^\dagger_{}\xi\phi^\dagger_{}\phi -(\mu_\Delta^2 +\lambda_{\xi\Delta}^{}\xi^\dagger_{}\xi+ \lambda_{\phi\Delta}^{}\phi^\dagger_{}\phi )\textrm{Tr}\left(\Delta^\dagger_{}\Delta\right) \nonumber\\
&&- \sum_{a=1}^{n}\left(M_{a}^{2}\sigma^\dagger_a \sigma_a^{}  + \kappa_a^{} \sigma_a^{} \phi^T_{} i \tau_2^{} \Delta \phi + \mu_a^{} \sigma_a^{} \xi^2_{} \right)
  \nonumber\\
&&-\frac{1}{2} f \bar{l}_{L}^c i \tau_2^{} \Delta l_{L}^{} - \sum_{i=1,2}^{}y_{i3}^{} \xi \bar{\nu}_{Ri}^{} \nu_{R3}^{c} +\textrm{H.c.}\,.
\end{eqnarray}
Note one of the four couplings $\kappa_{a,b\neq a}^{}$ and $\mu_{a,b\neq b}^{}$ can always keep complex after taking any phase rotations.

\section{Neutrino mass}

When the Higgs singlet $\xi$ develops its VEV $\langle\xi\rangle$ for the $U(1)_{B-L}^{}$ symmetry breaking, the heavy Higgs singlets $\sigma_a^{}$ can have a suppressed VEV,
\begin{eqnarray}
\langle\sigma_a^{}\rangle \simeq - \frac{\mu_a^{\ast}\langle\xi\rangle^2_{}}{M_{a}^2}\ll \langle \xi\rangle ~~\textrm{for}~~M_a^{}\gtrsim \mu_a^{}\,,~~M_a^{}\gg\langle\xi\rangle\,.
\end{eqnarray}
We hence can obtain a cubic coupling between the Higgs triplet $\Delta$ and the SM Higgs doublet $\phi$, i.e.
\begin{eqnarray}
\mathcal{L}\supset - \rho  \phi^T_{} i \tau_2^{} \Delta \phi +\textrm{H.c.} ~\textrm{with}~ \rho=\sum_a^{}\rho_a^{}=\sum_{a}^{} \kappa_a^{} \langle \sigma_a^{}\rangle\,.
\end{eqnarray}
From this triplet-doublet coupling, the Higgs triplet $\Delta$ can pick up a small VEV $\langle\Delta\rangle$ after the Higgs doublet $\phi$ acqures its VEV $\langle\phi\rangle\simeq 174\,\textrm{GeV}$ for the electroweak symmetry breaking, 
\begin{eqnarray}
\langle\Delta\rangle \simeq - \frac{\rho_{}^{\ast}\langle\phi\rangle^2_{}}{M_{\Delta}^2}\ll \langle \phi\rangle ~~\textrm{for}~~M_{\Delta}^{}\gg \rho\,,~~M_a^{}\gtrsim \langle\phi\rangle\,.
\end{eqnarray}
Here the Higgs triplet mass $M_\Delta^{}$ is dominated by 
\begin{eqnarray}
M_\Delta^2\simeq \mu_\Delta^2 +\lambda_{\xi\Delta}^{}\langle \xi \rangle^2_{} + \lambda_{\phi\Delta}^{} \langle \phi \rangle^2_{} \,.
\end{eqnarray}

\begin{figure}
\vspace{8cm} \epsfig{file=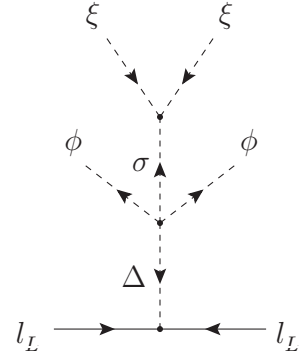, bbllx=5cm, bblly=6.0cm,
bburx=15cm, bbury=16cm, width=8cm, height=8cm, angle=0,
clip=0} \vspace{-10cm} \caption{\label{numass} The neutrino mass generation.}
\end{figure}

In the presence of the Higgs triplet VEV $\langle\Delta\rangle$, the left-handed neutrinos $\nu_L^{}$ can obtain their Majorana masses, 
\begin{eqnarray}
\mathcal{L}\supset -\frac{1}{2}m_\nu^{}\bar{\nu}_L^c \nu_L^{}+\textrm{H.c.}~~\textrm{with}~~m_\nu^{}=f \langle\Delta\rangle\,.
\end{eqnarray} 
If the VEV $\langle\Delta\rangle$ is at the eV scale, we can take the Yukawa couplings $f$ to be sizable and then yield the Majorana neutrino masses $m_\nu^{}$ at a desired level. The diagram for the neutrino mass generation is shown in Fig. \ref{numass}.

For a numerical example, we first fix
\begin{eqnarray}
 \langle \xi \rangle = \mathcal{O}(10\,\textrm{TeV})\,,
\end{eqnarray}
and then input
\begin{eqnarray}
&&M_{a}^{}= \mathcal{O}\left(10^{14}_{}\,\textrm{GeV}\right)\,,~~\mu_a^{} = \mathcal{O}\left(10^{13}_{}\,\textrm{GeV}\right)\,,\nonumber\\
&&\kappa_a^{}=\mathcal{O}(0.1)\,.
\end{eqnarray}
Consequently, we can have 
\begin{eqnarray}
\rho_a^{}=\mathcal{O}\left(10\,\textrm{eV}\right) \,.
\end{eqnarray}
By further inputting 
\begin{eqnarray}
M_\Delta= \mathcal{O}\left(\textrm{TeV}\right)\,,
\end{eqnarray}
we can obtain
\begin{eqnarray}
\langle\Delta\rangle= \mathcal{O}\left(0.1\textrm{eV}\right)\,,
\end{eqnarray}
and hence 
\begin{eqnarray}
m_\nu^{}= \mathcal{O}\left(0.01-0.1\textrm{eV}\right) ~~\textrm{for}~~f=\mathcal{O}(0.1-1)\,.
\end{eqnarray}

The generation of the small Higgs triplet VEV $\langle\Delta\rangle$ and hence the tiny neutrino masses $m_\nu^{}$ is the so-called type-II seesaw. For inducing the small VEV $\langle\Delta\rangle$, the Higgs triplet $\Delta$ should have a suppressed cubic coupling $\rho$ with the SM Higgs doublet $\phi$ if the Higgs triplet $\Delta$ has a mass $M_\Delta^{}$ at TeV scale. In the present work, the triplet-doublet coupling $\rho$ can be naturally achieved in an additional seesaw way, where the heavy Higgs singlets $\sigma_a^{}$ are far above the $U(1)_{B-L}^{}$ gauge symmetry breaking scale $\langle\xi\rangle$. Therefore, we may name this two-step seesaw mechanism as a double type-II seesaw \cite{ghsz2009}.

\section{Baryon asymmetry}

\begin{figure*}
\vspace{7.5cm} \epsfig{file=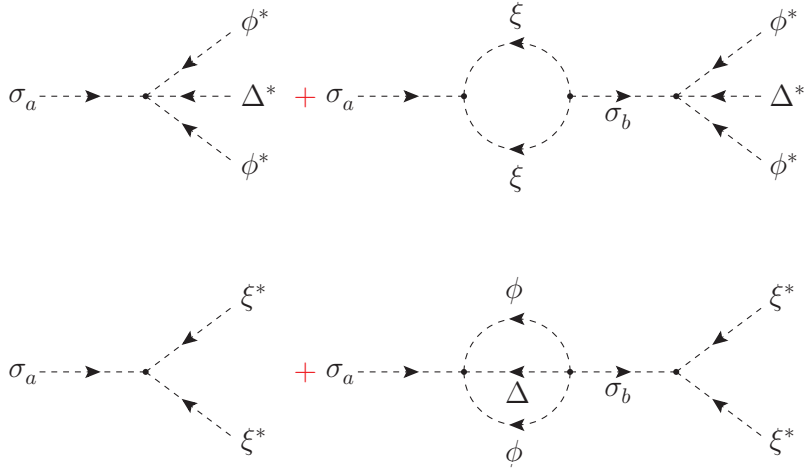, bbllx=5cm, bblly=6.0cm,
bburx=15cm, bbury=16cm, width=8cm, height=8cm, angle=0,
clip=0} \vspace{-8.5cm} \caption{\label{sdecay} The heavy Higgs singlet decays.}
\end{figure*}

As shown in Fig. \ref{sdecay}, the heavy Higgs singlets $\sigma_a^{}$ have two decay modes,
\begin{eqnarray}
\sigma_a^{}\rightarrow \phi^\ast_{}\phi^\ast_{}\Delta^\ast_{}\,,~~ \sigma_a^{} \rightarrow \xi^\ast_{}\xi^\ast_{}\,.
\end{eqnarray}
As long as the CP is not conserved, we can expect a CP asymmetry in the above decays, 
\begin{eqnarray}
\varepsilon_a^{} &=& 2\frac{ \Gamma(\sigma_a^{} \rightarrow \phi^\ast_{} \phi^\ast_{}\Delta^\ast_{}) - \Gamma(\sigma_a^{\ast} \rightarrow \phi\phi \Delta) }{\Gamma_a^{}}\nonumber\\
&=&- 2\frac{  \Gamma(\sigma_a^{}\rightarrow \xi^\ast_{}\xi^\ast_{}) - \Gamma(\sigma_a^{\ast}\rightarrow \xi \xi ) }{\Gamma_a^{}}\neq 0\,,\end{eqnarray}
where $\Gamma_a^{} $ is the total decay width,
\begin{eqnarray}
\Gamma_a^{} &=& \Gamma(\sigma_a^{}\rightarrow \phi^\ast_{}\phi^\ast_{}\Delta^\ast_{}) + \Gamma(\sigma_a^{} \rightarrow \xi^\ast_{}\xi^\ast_{}) \nonumber\\
&=&  \Gamma(\sigma_a^{\ast}\rightarrow\phi\phi\Delta) + \Gamma(\sigma_a^{\ast} \rightarrow \xi \xi) \,.
\end{eqnarray}
We can calculate the decay width at tree level and the CP asymmetry at one-loop level, 
\begin{eqnarray}
\Gamma_a^{} &=&\frac{1}{8\pi}\left( \frac{|\mu_a^{}|^2_{}}{M_a^2}+\frac{3}{32\pi^2_{}}|\kappa_a^{}|^2_{}  \right) M_{a}^{}\,,\\
\varepsilon_a^{} &=&-\frac{3}{32\pi^3_{}} \sum_{b\neq a}^{}\frac{\textrm{Im}\left(\kappa^\ast_{a}\kappa_b^{}\mu_a^{}\mu_b^\ast\right)}{\frac{|\mu_a^{}|^2}{M_{a}^2}+\frac{3|\kappa_a^{}|^2_{}}{32\pi^2_{}}}\frac{1}{M_b^2-M_a^2}\nonumber\\
&=& -\frac{3}{32\pi^3_{}}\sum_{b\neq a}^{}\frac{|\kappa^{}_{a}\kappa_b^{}\mu_a^{}\mu_b^{}|\sin\delta_{ab}^{}}{\frac{|\mu_a^{}|^2}{M_{a}^2}+\frac{3|\kappa_a^{}|^2_{}}{32\pi^2_{}}}\frac{1}{M_b^2-M_a^2}\,.
\end{eqnarray}
Here $\delta_{ab}^{}$ is the relative phase among the parameters $\rho_{a,b}^{}$ and $\kappa_{a,b}^{}$.

After the heavy Higgs singlets $\sigma_a^{}$ go out of equilibrium, their decays can generate a lepton asymmetry $L_\Delta^{}$ stored in the Higgs triplet $\Delta$. For demonstration, we simply assume the heavy Higgs singlet $\sigma_1^{}$ much lighter than the other heavy Higgs singlets $\sigma_{2,...}^{}$. In this case, the $\sigma_1^{}$ decays should dominate the final $L_\Delta^{}$ asymmetry, i.e.
\begin{eqnarray}
L_\Delta^{}&=&\varepsilon_1^{}\left(\frac{n^{eq}_{\sigma_1} }{s}\right)\left|_{T=T_D^{}}^{}\right.\,,
\end{eqnarray}
where the symbols $n^{eq}_{\sigma_1}$ and $T_D^{}$ respectively are the equilibrium number density and the decoupled temperature of the heavy Higgs singlets $\sigma_1^{}$, while the character $s$ is the entropy density of the universe \cite{kt1990}.

The cubic coupling between the Higgs triplet and doublet can go into equilibrium at a temperature much below the electroweak scale even if it appears before the electroweak symmetry breaking. Actually, we can estimate the interaction rate of the triplet-doublet coupling, 
\begin{eqnarray}
\Gamma_{\phi\phi\Delta}^{}\sim \left\{\begin{array}{cc}\frac{|\rho|^2_{}}{T} &\textrm{for}~~T>M_\Delta^{}\,,\\
[2mm]
\frac{|\rho|^2_{}}{M_\Delta} &\textrm{for}~~T<M_\Delta^{}\,,\end{array} 
\right.
\end{eqnarray}
and then require it to keep smaller than the Hubble constant $H(T)$ above the electroweak scale, i.e.
\begin{eqnarray}
\Gamma_{\phi\phi\Delta}^{}< H(T)\sim \frac{T^2_{}}{10^{19}_{}\,\textrm{GeV}}~~\textrm{for}~~T\gtrsim 100\,\textrm{GeV}\,.
\end{eqnarray}
The above condition can be achieved for $|\rho|\lesssim \mathcal{O}(\textrm{keV})$ and $M_\Delta^{}=\mathcal{O}(\textrm{TeV})$. Therefore, the lepton asymmetry $L_\Delta^{}$ stored in the Higgs triplet $\Delta$ will become a lepton asymmetry stored in the SM lepton doublets $l_L^{}$ before the electroweak symmetry breaking, thanks to the fast decays of the Higgs triplet into the lepton doublets. The sphaleron processes then can partially transfer this lepton asymmetry to a baryon asymmetry \cite{ht1990}, 
\begin{eqnarray}
B=-\frac{28}{79} L_\Delta^{}\,.
\end{eqnarray}

As a numerical example, we take
\begin{eqnarray}
\!\!&&M_{1}^{}= 10^{14}_{}\,\textrm{GeV}\,,~~|\mu_1^{} |= 3\times 10^{12}_{}\,\textrm{GeV}\,,~~|\kappa_1^{}|=0.3\,;\nonumber\\
\!\!&&M_{2}^{}= 10^{15}_{}\,\textrm{GeV}\,,~~|\mu_2^{} |=3\times 10^{13}_{}\,\textrm{GeV}\,,~~|\kappa_2^{}|=0.3\,.\nonumber\\
\!\!&&
\end{eqnarray}
The CP asymmetry $\varepsilon_1^{}$ then should be 
\begin{eqnarray}
\varepsilon_1^{}=-1.4\times 10^{-5}_{}\delta_{12}^{}\,.
\end{eqnarray}
Meanwhile, the weak washout condition can be satisfied \cite{kt1990},
\begin{eqnarray}
\!\!\!\!\Gamma_1^{}=7\times 10^9_{}\,\textrm{GeV}  <H(T)\left|_{T=M_1^{}}^{}\right. =1.5\times 10^{10}_{}\,\textrm{GeV} \,.
\end{eqnarray}
Here $H(T)$ is the Hubble constant,
\begin{eqnarray}
H(T)=\left(\frac{8\pi^{3}_{}g_{\ast}^{}}{90}\right)^{\frac{1}{2}}_{}\frac{T^2_{}}{M_{\textrm{Pl}}^{}}\,,
\end{eqnarray}
with $M_{\textrm{Pl}}^{}\simeq 1.22\times 10^{19}_{}\,\textrm{GeV}$ being the Planck mass and $g_{\ast}^{}=122$ being the relativistic degrees of freedom (the SM fields plus the three right-handed neutrinos $\nu_{R}^{}$, the Higgs triplet $\Delta$, the Higgs singlet $\xi$ and the $U(1)_{B-L}^{}$ gauge field.). The baryon number hence can be approximately given by \cite{kt1990}
\begin{eqnarray}
B =-\frac{28}{79} L_\Delta^{} \sim  -\frac{28}{79} \frac{\varepsilon_{1^{}}^{}}{g_\ast^{}} =10^{-10}_{}\left(\frac{\sin\delta}{2.4\times 10^{-3}_{}}\right)\,.
\end{eqnarray}

\section{Dark fermions}

As shown in Eq. (\ref{lag}), the three right-handed neutrinos $\nu_{R}^{}$ have two Yukawa couplings with the $U(1)_{B-L}^{}$ Higgs singlet $\xi$, i.e.
\begin{eqnarray}
\mathcal{L} &\supset& -\left[\bar{\nu}_{R1}^{} ~\bar{\nu}_{R2}^{}\right]\left[\begin{array}{cc}y_{13}^{}\\
[2mm]
y_{23}^{}\end{array}\right] \nu_{R3}^{c}+\textrm{H.c.} \nonumber\\
&=& - \left[\bar{\zeta}_{R}^{}~~\bar{\chi}_R^{}\right] \left[\begin{array}{cc}0\\
[1mm]y_{\chi}\end{array}\right] \nu_{R3}^{c}+\textrm{H.c.}~~\textrm{with}\nonumber\\
&&\zeta_R^{}=\frac{y_{23}^{}\nu_{R1}^{}-y_{13}^{}\nu_{R2}^{}}{\sqrt{y_{13}^2+y_{23}^2}}\,,\nonumber\\
&&\chi_R^{}= \frac{y_{13}^{}\nu_{R1}^{}+y_{23}^{}\nu_{R2}^{}}{\sqrt{y_{13}^2+y_{23}^2}}\,,\nonumber\\ 
&&y_\chi^{}=\sqrt{y_{13}^2+y_{23}^2}\,.
\end{eqnarray}
So, the third right-handed neutrino $\nu_{R3}^{}$ and the linear combination $\chi_R^{}$ of the two right-handed neutrinos $\nu_{R1,2}^{}$ can form a Dirac particle after the $U(1)_{B-L}^{}$ symmetry breaking, i.e.
\begin{eqnarray}
\mathcal{L} &\supset&  i \bar{\chi}\gamma^\mu_{}\partial_\mu^{} \chi - m_\chi^{} \bar{\chi}\chi \nonumber\\
&& \textrm{with}~~ \chi = \chi_{R}^{}+\nu_{R3}^{c}\,,~~ m_\chi^{}= y_{\chi}^{} \langle\xi\rangle\,.
 \end{eqnarray}
Meanwhile, the other linear combination $\zeta_R^{}$ of the two right-handed neutrinos $\nu_{R1,2}^{}$ has no Yukawa couplings so that it should be a massless state.

Since the Dirac fermion $\chi$ is stable, it can be expected to serve as a dark matter particle. The dark matter annihilation and scattering could be determined by the gauge interactions,
\begin{eqnarray}
\mathcal{L} &\supset &g_{B-L}^{} Z_{B-L}^\mu \left\{\sum_{i=1}^{3}\left(\frac{1}{3}\bar{d}_i^{}\gamma_\mu^{} d_i^{} + \frac{1}{3}\bar{u}_{i}^{}\gamma_\mu^{}u_{i}^{}-\bar{e}_{i}^{}\gamma_\mu^{} e_{i}^{}  \right.\right.\nonumber\\
&&\left.-\bar{\nu}_{Li}^{}\gamma_\mu^{}\nu_{Li}^{}\right)-4 \bar{\zeta}_{R}^{}\gamma_\mu^{}\zeta_{R}^{}  -  i 2\textrm{Tr}\left[\left(\partial_\mu^{}\Delta\right)^\dagger_{} \Delta-\textrm{H.c.} \right] \nonumber\\
&&\left.  -\frac{1}{2} \bar{\chi}\gamma_\mu^{}\left(9-\gamma_5^{}\right)\chi \right\}\,.
\end{eqnarray}
The gauge coupling $g_{B-L}^{}$ then should have an upper bound from the perturbation requirement, i.e.
\begin{eqnarray}
\label{gbl}
\frac{9}{2}g_{B-L}^{} < \sqrt{4\pi} \Rightarrow g_{B-L}^{}<  \frac{\sqrt{16\,\pi}}{9}\,,
\end{eqnarray} 
while the gauge boson mass $M_{Z_{B-L}}^{}$ should be
\begin{eqnarray}
M_{Z_{B-L}^{}}^{}&= & \sqrt{2} g_{B-L}^{} \langle \xi\rangle \,.
\end{eqnarray}
Currently the experimental constraints on the $U(1)_{B-L}^{}$ symmetry breaking is \cite{afpr2017},
\begin{eqnarray}
\label{low}
\frac{M_{Z_{B-L}^{}}^{}}{g_{B-L}^{} }  \gtrsim 7 \,\textrm{TeV} \Rightarrow \langle \xi \rangle \gtrsim 5\,\textrm{TeV}\,.
\end{eqnarray}

The thermally averaging dark matter annihilating cross section can be computed by \cite{bhkk2009}
\begin{eqnarray}
\label{ann}
\langle\sigma_{\textrm{A}}^{} v_{\textrm{rel}}^{} \rangle&=& \sum_{f=d,u,e,\nu_{L}^{},\zeta_R^{}}^{}\langle\sigma(\chi+\chi^c_{}\rightarrow f+f^c_{}) v_{\textrm{rel}}^{}\rangle \nonumber\\
&&+ \langle\sigma(\chi+\chi^c_{}\rightarrow \Delta+\Delta^\ast_{}) v_{\textrm{rel}}^{}\rangle \nonumber\\
&\simeq &\frac{2835 g_{B-L}^4}{8\pi} \frac{m_\chi^2}{M_{Z_{B-L}^{}}^4} \nonumber\\
&=& \frac{2835}{32\pi} \frac{m_\chi^2}{\langle\xi\rangle^4_{}} =  \frac{2835}{32\pi} \frac{y_{\chi}^2}{\langle\xi\rangle^2_{}}  \,.
\end{eqnarray}
The dark matter relic density then can well approximate to 
\begin{eqnarray}
\label{relic}
\Omega_\chi^{} h^2 \simeq \frac{0.1\,\textrm{pb}}{\langle \sigma_{\textrm{A}}^{} v_{\textrm{rel}}^{} \rangle} &=&0.1\,\textrm{pb}\times \frac{32\pi \langle\xi\rangle^4_{}}{2835 m_{\chi}^2}\nonumber\\
&=&0.1\,\textrm{pb}\times \frac{32\pi \langle\xi\rangle^2_{}}{2835 y_{\chi}^2} \,.
\end{eqnarray}
It should be noted that Eqs. (\ref{ann}) and (\ref{relic}) are based on the assumption, 
\begin{eqnarray}
\label{y34}
4m_\chi^2 \ll M_{Z_{B-L}^{}}^2 \Rightarrow y_{\chi}^{2}\ll \frac{1}{2} g_{B-L}^{2} < \frac{8\pi}{81}\Rightarrow y_\chi^{}< \frac{\sqrt{8\pi}}{9}\,.\nonumber\\
&&
\end{eqnarray}

By inserting the upper bound (\ref{y34}) into Eq. (\ref{relic}), we can put a constraint on the the VEV $\langle\xi\rangle$, i.e.
\begin{eqnarray}
\langle\xi\rangle &\simeq& \left(\frac{2835 y_{\chi}^2 \, \Omega_\chi^{} h^2}{32\pi \times  0.1\,\textrm{pb}} \right)^{\frac{1}{2}}_{}\nonumber\\
&=& 61 \,\textrm{TeV}\left(\frac{y_{\chi}^{}}{\sqrt{8\pi}/9}\right) \left(\frac{\Omega_\chi^{} h^2}{0.11}\right)^{\frac{1}{2}}_{}\nonumber\\
&<&61 \,\textrm{TeV}\left(\frac{\Omega_\chi^{} h^2}{0.11}\right)^{\frac{1}{2}}_{}\,,
\end{eqnarray}
besides the experimental limit (\ref{low}). The dark matter mass, 
\begin{eqnarray}
\label{dm1}
m_\chi^{} &\simeq &\left(0.1\,\textrm{pb}\times \frac{32\pi\langle\xi\rangle^4_{}}{2835 \Omega_\chi^{} h^2} \right)^{\frac{1}{2}}_{} \nonumber\\
&=& 4\,\textrm{TeV}\left(\frac{\langle \xi\rangle }{21\,\textrm{TeV}}\right)^2_{}\left(\frac{0.11}{\Omega_\chi^{} h^2}\right)^{\frac{1}{2}}_{} \,,
\end{eqnarray}
thus should be in the range, 
\begin{eqnarray}
\label{dm2}
&&227\,\textrm{GeV}\left(\frac{0.11}{\Omega_\chi^{} h^2}\right)^{\frac{1}{2}}_{}\lesssim
m_\chi^{} < 34\,\textrm{TeV}\left(\frac{0.11}{\Omega_\chi^{} h^2}\right)^{\frac{1}{2}}_{}\nonumber\\
&&\textrm{for}~~5\,\textrm{TeV} \lesssim \langle\xi\rangle < 61\,\textrm{TeV}\,.
\end{eqnarray}

The gauge interactions can also mediate the dark matter scattering off nucleons. The dominant spin-independent cross section is \cite{jkg1996}
\begin{eqnarray}
\sigma_{\chi N}^{}&=& \frac{81g_{B-L}^4}{4 \pi} \frac{\mu_r^2 }{M_{Z_{B-L}^{}}^4} \nonumber\\
&=& \frac{81 }{16\pi} \frac{\mu_r^2 }{\langle\xi\rangle^4_{}}=   \frac{2 }{35} \frac{\mu_r^2 }{m_\chi^2} \frac{0.1\,\textrm{pb}}{\Omega_\chi^{} h^2_{} }\,.
\end{eqnarray}
Here $\mu_r^{}=m_N^{}m_\chi^{}/(m_N^{}+m_\chi^{})$ is a reduced mass with $m_N^{}$ being the nucleon mass. As the dark matter is much heavier than the nucleon, we can simply read 
\begin{eqnarray}
\sigma_{\chi N}^{}&=& 2.9\times 10^{-45}_{}\,\textrm{cm}^2_{}\left(\frac{\mu_r^{}}{940\,\textrm{MeV}}\right)^2_{} \left(\frac{0.11}{\Omega_\chi^{} h^2_{}}\right)\nonumber\\
&&\times \left( \frac{4\,\textrm{TeV}}{m_\chi^{}}\right)^2_{} \,.
\end{eqnarray}
To match the dark matter direct detection results \cite{cui2017,aprile2018} into account, the dark matter mass should have a more stringent low limit,
\begin{eqnarray}
\label{dm3}
m_\chi^{} \gtrsim 4\,\textrm{TeV}\,.
\end{eqnarray}
So, the dark matter mass range (\ref{dm2}) should be modified by
\begin{eqnarray}
\label{range}
&&4\,\textrm{TeV}\left(\frac{0.11}{\Omega_\chi^{} h^2}\right)^{\frac{1}{2}}_{}\lesssim
m_\chi^{} < 34\,\textrm{TeV}\left(\frac{0.11}{\Omega_\chi^{} h^2}\right)^{\frac{1}{2}}_{}\nonumber\\
&&\textrm{for}~~21\,\textrm{TeV} \lesssim \langle\xi\rangle < 61\,\textrm{TeV}\,.
\end{eqnarray}

We also check if the massless $\zeta_R^{}$ fermion can decouple above the QCD scale to satisfy the BBN constraint on the effective neutrino number. For this purpose, we need consider the annihilations of the $\zeta_R^{}$ fermion into the relativistic species at the QCD scale,   
\begin{eqnarray}
\sigma_{\zeta}^{} &=&\sum_{f=d,u,s,e,\mu,\nu_L^{}}^{}\sigma(\zeta_R^{}+\zeta_R^c\rightarrow f+f^c) \nonumber\\
&=& \frac{6g_{B-L}^4}{\pi}\frac{s}{M_{Z_{B-L}}^4} =  \frac{3}{2\pi}\frac{s}{\langle\xi\rangle^4_{}}\,,
\end{eqnarray}
with $s$ being the Mandelstam variable. The interaction rate then should be \cite{gnrrs2003}
\begin{eqnarray}
\Gamma_{\zeta}^{} &=&\frac{\frac{T}{32\pi^4_{}}\int^{\infty}_{0} s^{3/2}_{} K_1^{}\left(\frac{\sqrt{s}}{T}\right) \sigma_{\zeta}^{}ds }{\frac{2}{\pi^2_{}}T^3_{}}= \frac{18}{\pi^3_{}} \frac{T^5_{}}{\langle\xi\rangle^4_{}}\,,
\end{eqnarray}
with $K_1^{}$ being a Bessel function. We take $g_\ast^{}(300\,\textrm{MeV})\simeq 61.75$ and then find 
\begin{eqnarray}
\left[\Gamma_{\zeta}^{} < H(T)\right]_{T\gtrsim 300\,\textrm{MeV}}^{}~~\textrm{for}~~\langle\xi\rangle \gtrsim 11\,\textrm{TeV}\,.
\end{eqnarray}
So, the massless $\zeta_R^{}$ fermion should be harmless for the parameter choice (\ref{range}).

\section{Conclusion}

In this paper we have shown a $U(1)_{B-L}^{}$ gauge symmetry can provide the lepton number violation for the Majorana neutrino mass generation, meanwhile, can predict the existence and guarantee the stability of the dark matter. Specifically, because of their Yukawa couplings to the Higgs singlet for spontaneously breaking the $U(1)_{B-L}^{}$ symmetry, three right-handed neutrinos without any Yukawa couplings to the SM can form a Dirac fermion besides a massless state. The massive right-handed neutrinos can serve as a stable dark matter particle while the massless one decouples safely. On the other hand, after this $U(1)_{B-L}^{}$ symmetry breaking, two or more heavy Higgs singlets can acquire their small VEVs to induce the cubic coupling between the type-II seesaw Higgs triplet and the SM Higgs doublet. Therefore, the left-handed neutrinos can naturally obtain their tiny Majorana masses even if the Higgs triplet is at the TeV scale. Through the interactions for generating the neutrino masses, the heavy Higgs singlets can decay to produce a lepton asymmetry stored in the Higgs triplet. The sphaleron processes then can partially transfer this lepton asymmetry to a baryon asymmetry.

\textbf{Acknowledgement}: This work was supported by the National Natural Science Foundation of China under Grant No. 11675100 and the Recruitment Program for Young Professionals under Grant No. 15Z127060004.

\end{document}